\documentclass{pasj00}
\usepackage{ulem}

\begin{document}
\SetRunningHead{K.\ Umetsu et al.}{Discovery of a strongly lensed galaxy at $z=3.9$}
\Received{}
\Accepted{}

\title
{Discovery of a Strongly Lensed Galaxy at $z=3.9$ behind a $z=0.83$
Galaxy Cluster
\thanks
{Based in part on data collected at the Subaru Telescope, which is
 operated by the National Astronomical Observatory of Japan.}}

\author{
   Keiichi \textsc{Umetsu}     \altaffilmark{1},
   Masayuki \textsc{Tanaka}    \altaffilmark{2},
   Tadayuki \textsc{Kodama}    \altaffilmark{3,5},\\
   Ichi \textsc{Tanaka}        \altaffilmark{4},
   Toshifumi \textsc{Futamase}        \altaffilmark{4},
   Nobunari \textsc{Kashikawa} \altaffilmark{3},
   Takako \textsc{Hoshi}       \altaffilmark{3,6}}

\altaffiltext{1}
 {Institute of Astronomy and Astrophysics, Academia Sinica, 
 P.O. Box 23--141,
 Taipei 106, Taiwan, Republic of China}
 \email{keiichi@asiaa.sinica.edu.tw}
\altaffiltext{2}
 {Department of Astronomy, School of Science, University of Tokyo, Tokyo 113--0033}
\altaffiltext{3}
 {National Astronomical Observatory of Japan, Mitaka, Tokyo 181--8588, Japan}
\altaffiltext{4}
 {Astronomical Institute, Tohoku University, Aoba-ku, Sendai 980--8578}
\altaffiltext{5}
 {European Southern Observatory, Karl-Schwarzschild-Str. 2, D-85748,
 Garching, Germany}
\altaffiltext{6}
 {Department of Physics, Meisei University, 2-1-1 Hodokubo, Hino, Tokyo 191-8605, Japan}

\maketitle

\begin{abstract}
We report the discovery and spectroscopic confirmation 
of three gravitationally-lensed images of a galaxy at $z=3.9$
in the background of a distant, rich cluster of galaxies at $z=0.83$,
on the basis of observations with Faint Object
Camera And Spectrograph (FOCAS) on the Subaru telescope.
We construct a simple lens model of the cluster mass distribution
based on Jee et al.'s  weak lensing mass estimates 
from deep, high-resolution images by Advanced Camera for Surveys (ACS)
 on the Hubble Space Telescope.
The lens model can account simultaneously for 
the observed image configuration
and the flux ratio of the closer pair located close to
the critical curve.
The parities of the three images are also 
consistent with the lensing hypothesis.
Since this galaxy is apparently bright ($i'_{\rm AB}\sim23.7$) for its
redshift due to the magnification, it serves as a good high redshift target
on which we can make extensive and detailed studies based on multi-wavelength
imaging and spectroscopy.
\end{abstract}

\section{Introduction}
\label{sec:intro}
Strong gravitational lensing due to the dense cores of clusters of galaxies
leads to the formation of 
multiple images and giant luminous arcs of 
background galaxies. 
The lensing observables such as image positions and flux-ratios
in turn provide strong constraints on the underlying gravitational
potential of the lensing clusters.
The strongly lensed images of background galaxies
have therefore been powerful observational tools to probe 
the mass distribution in cluster cores 
(Mellier et al. 1993; Kneib et al. 1993, 1996, 2003;
Hattori, Makino, \& Kneib 1999; Molikawa \& Hattori 1999;
 Broadhurst et al. 2005).

Exploring high-$z$ galaxies is extremely important for
understanding unsolved physical properties of forming galaxies
as well as understanding the
cosmic reionization history. The bottleneck is that such high-$z$
forming 
galaxies are too faint to be observed with present-day telescopes 
due to the $(1+z)^4$-dependence of the surface-brightness dimming
effect.
The most promising method is to utilize the gravitational magnification
by clusters as 
a biggest natural telescope -- Gravitational Lens Telescope
(Ellis et al. 2001; Hu et al. 2002; Kneib et al. 2004).
Furthermore,
detailed study of high redshift galaxies can only be made for strongly lensed
hence magnified galaxies (e.g., Pettini et al. 2002 on cB58). 

We have discovered a high-$z$ lensed galaxy
during the course of spectroscopic
follow-up of the RXJ0152.7$-$1357 cluster at $z=0.83$ 
as a part of our distant
cluster project on the Subaru called PISCES (Kodama et al. 2005).
In this paper, 
we present the spectroscopic and photometric properties of
the newly discovered triple lensed images
based on the observational data
taken with the ground-based Subaru telescope as well as on the data
with the 
high-resolution archival
Advanced Camera for Surveys (ACS)
on the Hubble Space
Telescope (HST).  A gravitational lens model of the cluster mass distribution
is developed for the three
lensed images.
Throughout the paper
we assume a $\Lambda$-dominated flat cosmology with
$\Omega_m=0.27,\Omega_{\Lambda}=0.73$,
and $h=0.7$, and we use magnitudes given in the AB system.

\section{Identification of the Lensed Objects}
\label{sec:obs}

\begin{figure}
\label{fig:spec}
  \FigureFile(80mm,80mm){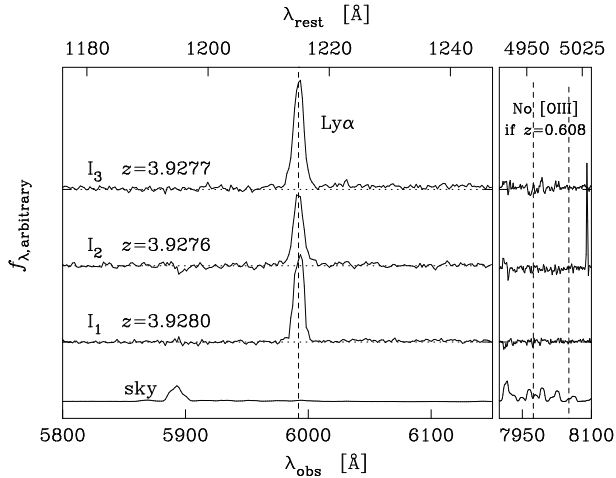}
\caption{
The optical 
spectra of three multiple  
images $(I_1,I_2,I_3)$ obtained with FOCAS on the Subaru telescope.
The zero-flux levels are indicated with horizontal dotted lines.
}
\label{fig:map}
\end{figure}

The lensed objects were first found as 
those having very similar
colors in $VRi'z'$ based on the optical imaging with Suprime-Cam
on the Subaru telescope
(see Kodama et al. 2005 for details of the optical data).
The photometric properties of the objects are summarized in Table 1.
All the three objects show strong continuum breaks in the $V-R$ color
($\sim 0.9)$
with a relatively flat SED at longer wavelength regime.
These photometric properties suggest that their redshifts are
high ($z\sim 4$) in analogy to Lyman break galaxies.

We have conducted spectroscopic follow-up observations during
11--14 October 2004 with FOCAS (Kashikawa et al. 2002) in 
the MOS mode.
We used a 300 lines $\rm mm^{-1}$ grating blazed at 
5500 $\rm \AA$
with 
the order-cut filter SY47.
The wavelength coverage was between 4700$\rm\AA$ and 9400 $\rm\AA$ with a pixel
resolution of $1.40\rm \AA\  pixel^{-1}$.
A slit width was set to $0\farcs 8$, which gave a resolution of
$\lambda/\Delta\lambda\sim500$.
We obtained four 1800s exposures 
for each object
nodded by $\sim0\farcs 6$ for each shot.
The object $I_2$ was slipped out of the slit for one shot
due to the nodding, and the net exposure on this object is $3\times1800$s.
Data reduction was performed in a standard manner using {\it IRAF}.
The objects $I_1$ and $I_2$ were very close to the spatial
edge of the slits,
and hence we will not discuss relative fluxes of 
these objects from the spectra.
Figure 1 shows
the obtained spectra of the 
the three images under concern.
All the three show a strong emission line around 
$\lambda_{\rm obs}=5992 {\rm \AA}$.
The emission line profile for each object was fitted with a Gaussian.
The central wavelengths of the lines are 5992.41$\rm\AA$, 
5991.95$\rm\AA$, and 5992.11$\rm\AA$ 
for $I_1$, $I_2$, and $I_3$, respectively, with a statistical error of
$\sim 0.3{\rm \AA}$.
Therefore, the objects lie at the same redshift within the errors.
The objects $I_1$ and $I_3$ show a clear continuum 
break shortward of this line,
although no such break is seen in $I_2$ due to the low {\it S/N}.
No other lines are observed in the spectra.
These objects are either (1) a gravitationally lensed 
  Ly$\alpha$ emitter at $z=3.928$ or (2) foreground [OII] emitters 
  at $z=0.608$.
The latter possibility is rejected, however.
We observe no [OIII] lines at the expected wavelength as shown in Fig.~1.
It is highly unlikely that galaxies show such a strong [OII] emission
without any detectable [OIII] emission.
The photometric properties of the objects are 
inconsistent with a $z=0.608$ galaxy either.
Their $V-R$ and $R-i'$ colors are too red for their $i'-z'$ colors to be
star forming galaxies at $z\sim0.6$.
We therefore conclude that these objects are 
Ly$\alpha$ emitters at $z=3.928$.
Note that 
the rms error in the redshifts is $\sigma(z)\sim 3\times 10^{-4}$.
In summary, these objects show very similar photometric properties and they lie
at exactly the same redshift --  these facts strongly suggest 
that they are originally
a single galaxy strongly lensed by the foreground cluster at $z=0.83$.

We also obtained the "on-the-fly" processed HST/ACS images of the field 
from the HST archive (HST GTO Proposal 9290: Ford, H.). The cluster was 
observed in a $2 \times 2$ mosaic pattern with an overlap of $\sim50\arcsec$ 
regions in F675W, F775W, and F850LP filters. 
The total exposure per pointing
is about 4.8 ks for each filter.  
%
We show in Fig.~2 the HST image
of the lensed objects 
and the Subaru image in Fig.~3.
The objects all apparently look like the edge-on disk, which is probably 
due to the effect of strong lensing.

\begin{table*} 
\label{tab:so}
\caption{
Photometric properties of 
three gravitationally lensed images from Subaru/Suprime-Cam and 
HST/ACS data$^*$}
\begin{center}
\begin{tabular}{lccccccccc}
\hline\hline
Object & R.A.    & Dec.    & $i'^{\dagger}$ &    
$V-R$ & $R-i'$ & $i'-z'$  & $r_h^{\ddagger}$\\
& (J2000) & (J2000) & (AB-mag) & (AB-mag) & (AB-mag) & (AB-mag) & ($''$)\\
\hline
$I_1$ & 01 52 45.32 & -13 57 08.6 & $23.70\pm0.04$ & 
$+0.85\pm 0.01 $& $-0.03\pm 0.01$ &$-0.07\pm 0.03$  &
0.141\\
$I_2$ & 01 52 45.14 & -13 57 04.9 & $24.05\pm 0.04$ & 
$+0.90\pm 0.02$  &  $+0.08\pm 0.02$ & $+0.18\pm 0.03$ &
0.125\\
$I_3$ & 01 52 44.41 & -13 56 56.7 & $23.89\pm0.04$ & 
$+0.88\pm 0.01$ &$-0.06\pm 0.02$  &$+0.02\pm 0.03$   &
0.127\\
\hline 
 \multicolumn{8}{l}{\scriptsize
 * The total magnitudes 'MAG\_AUTO' (SExtractor) are used for $i'_{\rm AB}$.
  Photometric apertures of $1\arcsec$ are used for color measurements.}\\
 \multicolumn{8}{l}{\scriptsize
 $\ddagger$ Half-light radii (FLUX\_RADIUS by SExtractor) measured in the HST/ACS
 F675W-band image}\\
\end{tabular}\\
\end{center}
\label{tab:sources}
\end{table*}

\begin{figure}
\label{fig:image}
\begin{center}
  \FigureFile(80mm,80mm){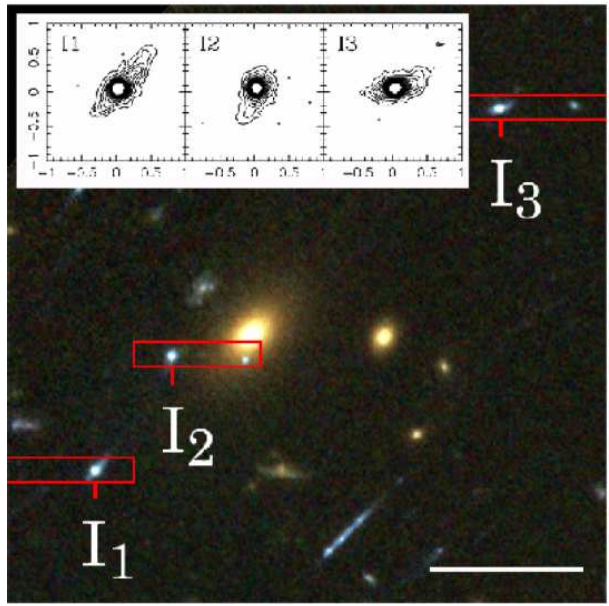}
\end{center}
\caption{ 
 Zoom-in view of 
a $20\arcsec \times 20\arcsec$
region around the three
gravitationally-lensed objects $(I_1,I_2,I_3)$ 
in the HST/ACS composite color image
constructed using F675W, F775W, and F850LP filters. 
The stick at the bottom right has a $5\arcsec$ length.
The red open 
rectangles indicate
the slit locations for the  
Subaru/FOCAS
multi object spectroscopy of 
$(I_1,I_2,I_3)$.
The inset plots show the surface brightness contours of
$(I_1,I_2,I_3)$ from the F675W-band data.
The coordinates are in arcseconds.
The lowest contour and the contour
interval are both at a $3\sigma$ rms noise level of 
$25.06$ mag arcsec$^{-2}$.
} 
\label{fig:images}
\end{figure}

\section{Strong Gravitational Lensing}
\label{sec:model}

\begin{figure}
\label{fig:gl}
\begin{center}
  \FigureFile(85mm,85mm){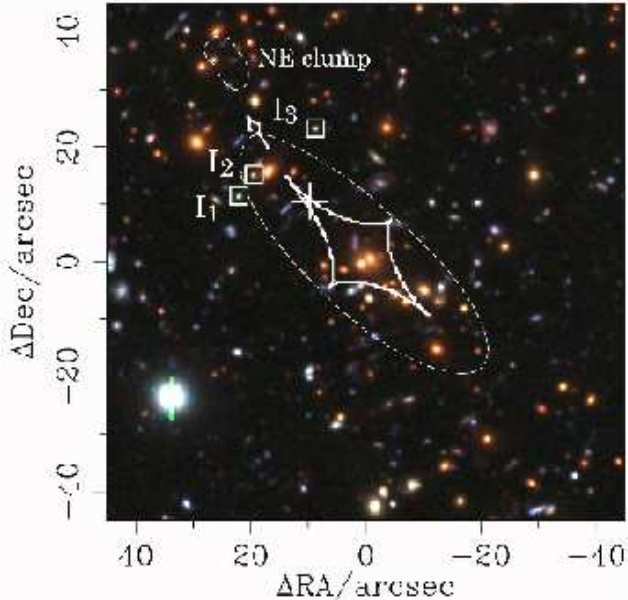}
\end{center}
\caption{
$VRi'$
color image of the 
 central $90\arcsec \times 90\arcsec$ region of RXJ0152 ($z=0.83$)
 obtained with Subaru/Suprime-Cam.
 North is top and East is left.
 Also overlayed are the caustic (thick solid lines) and the critical
  curves (thin dashed lines) predicted by our best fitting model.
 The cross represents the predicted position of the source galaxy at $z=3.9$.
 The open squares represent the predicted positions of the multiple
 images $(I_1,I_2,I_3)$.
}
\label{fig:map}
\end{figure}

\subsection{Lens Model}

The RXJ0152 gravitational lens system shows the colinear, three multiple
images $(I_1,I_2,I_3)$ of a source galaxy at $z=3.9$,
from which one expects that the lensed images could be associated with a cusp
caustic produced by an elliptical potential (e.g., Schneider, Ehlers, \&
Falco 1992).
In constructing a lens model of the cluster mass distribution,
we closely
follow weak lensing mass estimates by Jee et al. (2005) 
based on the HST/ACS observations with superb angular-resolution and 
photometric-depth. 
We adopt Jee et al.'s 
parameterized singular isothermal
ellipsoid model
(SIE: Kormann et al. 1994; King \& Schneider 2001) 
centered between
the two central brightest cluster galaxies,
for describing the elongated global mass distribution.
The SIE requires three model parameters: Einstein 
radius
$\theta_{\rm E}$, 
minor-to-major axis ratio  $f(= b/a)$, 
and orientation angle $\alpha$.
Jee et al. obtained the best-fit parameters of $\theta_{\rm E}=7\farcs
19 \pm 0\farcs 72$  ($ z_s = 1.3$)
and
$f=0.36\pm 0.1$. This Einstein angle corresponds to 
the 1-D velocity dispersion of $\sigma_v^{\rm SIE}=940\pm 168 {\rm km}s^{-1}$.
For the source redshift of  $z_s=3.9$, this angle is scaled to
$\theta_{\rm E}=15\farcs 4$. 
The cluster contains several mass clumps
associated with the cluster galaxy concentrations. 
As the three images are located in between the cluster center and the
North-East substructure (mass clump B of Jee et al.),
we take this clump into account
in our lens model in the following way:
We assume the clump mass of $M=3.4\times 10^{13}M_{\odot}$
estimated by Jee et al. is 
the virial mass of the clump.
We then convert the mass into the 1-D velocity dispersion
of an SIS halo ($\sigma_{v}^{\rm SIS}\propto M^{1/3}$: 
see Bartelmann, King, \& Schneider 2001).
Accordingly,
our model consists of one SIE as the global cluster halo
and one SIS as the North-East substructure. 
We treat the SIE parameters ($\theta_{\rm E}$, $f$, $\alpha$)
and the (un-lensed) source position 
$\mbox{\boldmath $\theta$}_S$
as free parameters  of our lens model.

We then constrain the 
5 model parameters via $\chi^2$-fitting to the
observation data: 
the positions and the fluxes of the three images.
In the model-fitting, we use as constraints 
the thee image positions 
$\mbox{\boldmath $\theta$}_{I_k} (k=1,2,3)$
relative to the
center of the SIE halo;
we further include the flux ratio $I_2/I_1$ of the closer pair
as it gives a strong constraint on the location of the critical curve.
The $\chi^2$ is then 
given as
\begin{equation}
\chi^2=\sum_{k=1}^{3}
\frac{ ( \mbox{\boldmath $\theta$}_{I_k}
        -\mbox{\boldmath $\theta$}^{\rm M}_{I_k} )^2 }
     { \sigma_{\rm pos}^2 }
+ 
\frac{ \left[ (I_2/I_1)-(I_2/I_1)^{\rm M}\right]^2 } 
     { \sigma_{I_2/I_1}^2 },
\end{equation}
where quantities with superscript M denote the model predictions,
$\sigma_{\rm pos}$ is
the positional
uncertainty of the multiple images, 
$\sigma_{\rm pos}\approx 0\farcs 1$,
and $\sigma_{I_2/I_1}$ is the rms error of the flux ratio $I_2/I_1$.
The image positions $\mbox{\boldmath $\theta$}_I$
of a source with the position $\mbox{\boldmath $\theta$}_S$
are obtained by solving the lens equation 
$  \mbox{\boldmath $\theta$}_S = \mbox{\boldmath $\theta$}_I - 
\mbox{\boldmath $\nabla$}
\psi( \mbox{\boldmath $\theta$}_I )$
(Schneider et al. 1992) with the effective lensing potential
$\psi=\psi_{\rm SIE}+\psi_{\rm SIS}$ in sum of an SIE and an SIS
potential.

\subsection{Model Predictions v.s. Observations}
We show in Fig.~3 
the derived best-fit model of 
the RXJ0152 lens system 
overlaied on the Subaru image.
 The minimum $\chi^2$ is 
 $\chi_{\rm min}^2/{\rm dof}=2.03/2$ for $2$ degrees of freedom (dof). 
 We quote the best-fit SIE parameters
 are $\theta_{\rm E}=14\farcs 3^{+0.4(+1.1)}_{-0.3(-0.7)}$,
$f=0.25^{+0.03(+0.08)}_{-0.03(-0.07)}$,
and $\alpha=47.\degree 7 \pm 0.2 (\pm 0.4)$ 
measured East of North,
with the $1\sigma$ ($2\sigma$) errors estimated from
$\Delta\chi^2\equiv \chi^2-\chi^2_{\rm min}=1$ (4) 
in the 5 parameter space.

The predicted source position is inside and close to a cusp caustic
produced by the main cluster potential as expected.
For such an image configuration, the image $I_2$ is predicted
to have opposite parity as $I_1$ and $I_3$. 
In particular, the closest images $(I_1,I_2)$ are mirror images of each
other
with respect to the critical curve (see Fig.~3). 
The high-resolution ACS images reveal more detailed structures of the
lensed images that were not spatially resolved in the Subaru images.
Figure 2 clearly shows that the observed image parities are consistent 
with the
lensing hypothesis: 
each image contains a bright core and extended
emission with lower surface brightness, and 
a clear mirror symmetry between $I_1$ and $I_2$
is seen in the HST/ACS data. 
 The predicted amplifications using the best-fit model are
 $\mu_1=7.2$, $\mu_2=5.8$,
 and $\mu_3=3.0$ for $I_1,I_2$, and $I_3$, respectively.
The total amplification factor over the three images 
is hence $\approx 16$.
 Accordingly, the predicted magnification ratios are
 $I_2/I_1=0.80$ and $I_3/I_1=0.42$,
whereas 
the flux ratios from the Subaru $i'$-band data
are $I_2/I_1=0.82\pm 0.04$ and  $I_3/I_1=0.89\pm 0.05$.
Our simple lens model thus reproduces correctly the observed flux ratio
of $I_2/I_1$ while it fails for $I_3/I_1$; more than a factor 2 smaller
than the observed flux ratio.
In this study, 
we will not assess
the significance of the difference between the observed and predicted
flux-ratios for $I_3/I_1$
as it requires more detailed lens modeling 
for constraining  all possible substructures (cf. Kneib et al. 1996).
It is expected that 
the image $I_3$ could be easily magnified by local perturbers 
since its angular size 
(see Table 1) is
sufficiently small compared with that of the lensing potential of
possible perturbers (Chiba 2002).

\section{Intrinsic Properties of the Source Galaxy}

The derived lens model allows us to
inspect the 
intrinsic properties of the un-lensed source 
from the observed images.
In what follows, we only refer to the images $I_1$ and $I_2$
for which our lens model can accurately reproduce the observations.
From the predicted
magnifications, 
the magnitude of the un-lensed source 
is 
estimated as $i'_{\rm AB}\approx 25.9$.
This corresponds to
a UV-luminosity of 
$F(1560{\rm \AA})=6.3 \times 10^{40}$
erg/s/$\rm \AA$ or $M(1560{\rm \AA})=-20.2$ mag.
From this, we estimate a star formation rate (SFR) 
of the source
galaxy to be 
$6 M_\odot {\rm yr}^{-1}$ without dust extinction
correction (Madau et al. 1998). 
 The observed Ly$\alpha$ flux indicates SFR$\sim 3M_\odot\ {\rm yr}^{-1}$
 (Kennicutt 1998; Brocklehurst 1971), 
which is lower than SFR(1560$\rm \AA$). This is suggestive of dust extinction.
Further, the superb HST/ACS angular resolution
together with our lens model gives us 
size information of the source galaxy. The half-light radii $r_h$
of the
lensed images are listed in Table 1 ({FLUX\_RADIUS} given by 
SExtractor: Bertin \& Arnouts 1996). Note that even for the ACS images
the FWHM of PSF is comparable to that of the lensed images.
Taking this into account, we give a constraint on the intrinsic source size
as 
$ r_h \lesssim 0\farcs 02 $ ($r_{3\sigma}\lesssim 0\farcs 08$; see
Fig.~2),
or 
$\lesssim 150$ pc ($\lesssim 600$ pc)
in physical scales, 
assuming a simple one-dimensional amplification
near the critical curve (see Ellis et al. 2001 for a similar case).

\section{Summary}
\label{sec:summary}

We report the discovery and spectroscopic confirmation of
three gravitationally lensed images of a distant galaxy 
in the background of the rich cluster RXJ0152.7$-$1357 ($z=0.83$).
The similarity of the photometric and spectroscopic properties of 
the three images from Subaru observations
provides strong evidence that they are
gravitationally lensed images of 
a Ly$\alpha$ emitter at $z=3.9$.
A lens model was constructed on the basis of  
Jee et al.'s weak lensing
mass estimates from 
the HST/ACS data.
Our simple lens model can
account simultaneously for the observed image configuration 
and the flux ratio of the closer pair located 
close to
the critical curve.
The image parities in the ACS image are also 
consistent with the lensing hypothesis.

 To further investigate the properties of this source galaxy at $z=3.9$,
  we need longer wavelength data, such as $K$-band which samples its
  rest-frame $B$-band light or {\it Spitzer/Astro-F} bands at 3--8$\mu$m
  which sample the light from underground old populations 
  related to the stellar mass of this galaxy.
A more detailed modeling of the lens mass distribution is important,
as a well-constrained cluster mass model in conjunction 
with different sets of multiple images with suitable redshifts 
serves as a powerful probe of the geometry of the universe
(Futamase \& Yoshida 2001; Golse, Kneib, \& Soucail 2002).

\section*{Acknowledgements}
We acknowledge M. Chiba, M. Takada, T. Chiueh,
and B.-C. Hsieh
for useful discussions.
We thank our referee, J.-P. Kneib, for 
his useful remarks helping to improve the paper.
This work is financially supported in part by a Grant-in-Aid for the
Scientific Research (No.\ 15740126)
by the Ministry of Education, Culture, Sports, Science and Technology.
M.T. acknowledges the JSPS research fellowship.
Some of the data presented in this paper were obtained 
from the Multimission Archive at the Space Telescope Science 
Institute (MAST).
STScI is operated by the Association of Universities 
for Research in Astronomy, Inc., under NASA contract NAS5-26555.

\end{document}